\documentclass[aps,prl,twocolumn,showpacs,floatfix,nofootinbib,%
groupedaddress]{revtex4-1}
\usepackage{epsfig,subfigure}
\usepackage{graphicx}
\usepackage{amsmath}
\usepackage{amssymb}
\usepackage{bm}
\usepackage{bbm}
\usepackage{multirow}
\begin{document}
\title{$\bm{pp\to J/\psi+\Upsilon+X}$
as a clean probe to the quarkonium production mechanism
}
\author{P. Ko}
\email[]{pko@kias.re.kr}
\author{Chaehyun Yu}
\email[]{chyu@kias.re.kr}
\affiliation{School of Physics, KIAS, Seoul 130-722, Korea}
\author{Jungil Lee}
\email[]{jungil@korea.ac.kr}
\affiliation{Department of Physics, Korea University, Seoul 136-701, Korea}
\begin{abstract}
We report that, unlike most inclusive quarkonium-production processes,
the production rate for $pp\to J/\psi+\Upsilon+X$ dominantly depends on two
not well-known long-distance nonrelativistic QCD matrix elements, 
$\langle O_8^{J/\psi}({}^3S_1)\rangle$ and
$\langle O_8^{\Upsilon}({}^3S_1)\rangle$ 
at leading order in the strong coupling constant, 
which account for the transition probabilities of the color-octet spin-triplet
heavy-quark-antiquark pairs $c\bar{c}_8({}^3S_1)\to J/\psi$ and
$b\bar{b}_8({}^3S_1)\to \Upsilon$, respectively. With the integrated luminosity
$\sim 100 \,\textrm{fb}^{-1}$ at the center-of-momentum energy
$\sqrt{s}=14$\,TeV we expect that approximately 1900
$pp\to J/\psi+\Upsilon+X$ events can be observed at the CERN Large Hadron
Collider by tagging muon pairs, which are enough to probe to the color-octet
mechanism. The forthcoming measurement may provide a useful constraint to
resolve the decade-old puzzle for the polarization of prompt $J/\psi$ at the
Fermilab Tevatron.
If corresponding measured rate is significantly less than the prediction,
it may imply that the current values for the color-octet matrix elements 
are overestimated.
\end{abstract}
\maketitle
As an effective field theory of QCD, 
the nonrelativistic QCD (NRQCD) factorization approach 
\cite{Bodwin:1994jh} has achieved great progress in understanding the
production and decay mechanism of heavy quarkonia \cite{Brambilla:2004wf}.
The factorization has been proved for the electromagnetic and light-hadronic
decays \cite{Bodwin:1994jh} and also for a few specific exclusive production
processes very recently \cite{Bodwin:2008nf,Bodwin:2009cb,Bodwin:2010fi}. The 
factorization conjecture for the inclusive production has been employed to
resolve the large surplus of $\psi(2S)$ and prompt $J/\psi$ at the Fermilab
Tevatron \cite{Braaten:1994vv}, which lead to a remarkable prediction that
the prompt $J/\psi$ must be transversely polarized at the large transverse
momentum ($p_T$) \cite{Braaten:1999qk}.
However, the empirical data for the polarization of prompt $J/\psi$
and $\psi(2S)$ measured 
by the CDF Collaboration \cite{Affolder:2000nn,Abulencia:2007us}
are in disagreement with these predictions. 
That is what we call the puzzle for the polarization of
prompt $J/\psi$ at the Tevatron.

These predictions strongly depend on the determination of the long-distance
NRQCD matrix element $\langle O_n^H ({}^{2s+1}L_J)\rangle$, which accounts
for the probability of the heavy-quark-antiquark pair $Q\bar{Q}$ with the
spectroscopic state ${}^{2s+1}L_J$ to evolve into a heavy quarkonium $H$. 
The color-singlet NRQCD matrix elements 
$\langle O_1^H({}^3S_1)\rangle$ for $H=J/\psi$ and $\Upsilon$ are well known
through precisely measured values for their leptonic decay rates. 
On the other hand, the color-octet counterparts such as 
$\langle O_8^H({}^3S_1)\rangle$, $\langle O_8^H({}^1S_0)\rangle$, and
$\langle O_8^H({}^3P_J)\rangle$ for $J=0$, 1, 2 are not determined accurately,
because in most of the hadroproduction rates for $H$ all of these
color-octet matrix elements involve simultaneously and the $p_T$ spectra are
not well distinguishable to be fit to the data \cite{Cho:1995ce,Braaten:2000cm}.
To make the situation worse, large next-to-leading-order corrections
dramatically modify the $p_T$ spectra of these
processes~\cite{Campbell:2007ws,Gong:2008sn,Gong:2008ft,Artoisenet:2008fc}.
As a result, the matrix elements determined from various processes are not
consistent with each other 
\cite{Gong:2008ft,Ko:1996xw,Klasen:2001cu,Artoisenet:2009xh,Chang:2009uj,%
Zhang:2009ym,Butenschoen:2009zy}.
In particular, 
the actual values for the color-octet matrix elements
$\langle O_8^{J/\psi}(^1S_0)\rangle$ and $\langle O_8^{J/\psi} (^3P_0) \rangle$
might be significantly smaller than the naive estimates based on the
velocity-scaling rules of NRQCD~\cite{Zhang:2009ym}.
In this situation, it would be highly desirable to find a hadroproduction
process that involves fewer numbers of color-octet channels.

In this paper, we propose that $pp\to J/\psi+\Upsilon+X$ at the CERN Large
Hadron Collider (LHC) could be a clean probe to the color-octet mechanism
of NRQCD and compute its $p_T$ spectrum and the total production rate 
at the center-of-momentum energy $\sqrt{s}=14$\,TeV. Unlike most
inclusive quarkonium-production processes~\cite{Braaten:1999qk,%
Qiao:2002rh,Li:2009ug,Qiao:2009kg}, the production rate for
$pp\to J/\psi+\Upsilon+X$ dominantly depends on two not well-known
long-distance factors $\langle O_8^{J/\psi}({}^3S_1)\rangle$ and
$\langle O_8^\Upsilon({}^3S_1)\rangle$ and 
the color-singlet contribution is suppressed at leading order (LO)
in the strong coupling constant $\alpha_s$ 
in the asymptotic limit $p_T\to\infty$.
The forthcoming measurement may provide a useful constraint to
resolve the decade-old puzzle for the polarization of prompt $J/\psi$ at the
Tevatron.

This paper is organized as follows. 
We first summarize the NRQCD factorization formula for 
$pp\to J/\psi+\Upsilon+X$ and describe a systematic way to determine
the dominant contributions. Next we provide detailed arguments to convince
the dominance of $\langle O_8^{J/\psi}({}^3S_1)\rangle$ and
$\langle O_8^\Upsilon({}^3S_1)\rangle$ contributions
in the $p_T$ spectrum at LO in
$\alpha_s$ and the typical velocity $v_Q$ of the heavy quark $Q$
in the quarkonium rest frame. Our prediction for the production rate
and $p_T$ spectrum at the LHC follows and
then we conclude.

The NRQCD factorization formula for the differential cross 
section $d \sigma$ of $pp\to J/\psi+\Upsilon+X$
has the following schematic form:
\begin{equation}
\label{xsec}%
d \sigma=
f_{i/p}\otimes f_{j/p}\otimes
d \hat{\sigma}_{ij\to c\bar{c}_{n_1}+b\bar{b}_{n_2}}
\langle O^{J/\psi}_{n_1} \rangle
\langle O^{\Upsilon}_{n_2} \rangle,
\end{equation}
where $f_{i/p}$ is the parton distribution function (PDF), $d\hat{\sigma}$
is the parton-level cross section, which is perturbatively calculable in
powers of $\alpha_s$, $\langle O^{H_i}_{n_i} \rangle$
is the NRQCD matrix element for the quarkonium $H_i$ involving the
NRQCD four-quark operator $O^{H_i}_{n_i}$ \cite{Bodwin:1994jh}. The subscript
$n_i$ represents the spectroscopic state of the $Q\bar{Q}$
pair. For example, $Q\bar{Q}_{n_i}=Q\bar{Q}_{n}({}^{2s+1}L_J)$ stands for
the color-singlet ($n=1$) or -octet ($n=8$) $Q\bar{Q}$ pair with the spin
$s$, the orbital angular momentum $L$, and the total angular momentum $J$
for $Q=c$ or $b$. 
The symbol $\otimes$ indicates the convolution over the partons' longitudinal
momentum fractions and the summation over partons $i$ and $j$ is assumed.
According to the velocity-scaling rules of NRQCD \cite{Bodwin:1994jh},
the production rates for the spin-triplet $S$-wave quarkonium through the
color-octet states $Q\bar{Q}_8({}^3S_1)$, $Q\bar{Q}_8({}^1S_0)$, and
$Q\bar{Q}_8({}^3P_J)$ for $J=0$, 1, 2 are suppressed by $v_Q^4$, $v_Q^3$,
and $v_Q^4$, respectively, compared to the color-singlet state
$Q\bar{Q}_1({}^3S_1)$. Let us call the product of these factors for a
process the ``velocity-scaling factor'' $\mathcal{V}$. Therefore,
Eq.~(\ref{xsec}) is a power series in $\alpha_s$, $v_c$, and $v_b$.

\begin{figure}
\epsfig{file=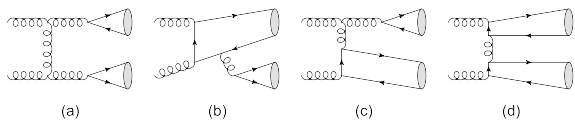,height=1.5cm}
\caption{\label{figpu}%
Typical Feynman diagrams for gluon-initiated parton processes for
$pp\to J/\psi+\Upsilon+X$ at order $\alpha_s^4$.
}
\end{figure}

\begin{figure}
\epsfig{file=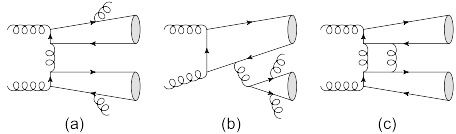,height=1.8cm}
\caption{\label{figjusing}%
Typical Feynman diagrams for the color-singlet contribution to
$pp\to J/\psi + \Upsilon +X$ at order $\alpha_s^6$, which is leading.
}
\end{figure}

At LO in $\alpha_s$, only $g g$ fusion and $q \bar{q}$
annihilation contribute to the parton processes 
$ij\to c\bar{c}_{n_1}+b\bar{b}_{n_2}$ of order $\alpha_s^4$ and the channels 
with $ij (ji)=gq$ and $g\bar{q}$ are missing. At the 
center-of-momentum energy of the LHC 
one probes the small-$x$ region of the PDF, where the gluon contribution
dominates over the quark contents. Therefore, we consider only the $gg$
initial states. Typical Feynman diagrams for the gluon-initiated parton
process $gg\to c\bar{c}_{n_1}+b\bar{b}_{n_2}$ are shown in Fig.~\ref{figpu},
where at least one $Q\bar{Q}$ pair is in a color-octet state. Leading
color-singlet contribution $c\bar{c}_1({}^3S_1)+b\bar{b}_1({}^3S_1)$
appears only at order $\alpha_s^6$, which are shown in Fig.~\ref{figjusing}.

We first consider the color-octet contribution
$c\bar{c}_8({}^3S_1)+b\bar{b}_8({}^3S_1)$ 
in Figs.~\ref{figpu} (a)--(d)  with $\mathcal{V}=v_c^4 v_b^4$, 
to which 36 Feynman diagrams contribute.
At large $p_T$ the double-fragmentation contribution
[Fig.~\ref{figpu}\,(a)] dominates because of the 
kinematic enhancement. Figure~\ref{figpu}\,(b) also represents mixed
contributions $c\bar{c}_8({}^3S_1)+b\bar{b}_1({}^3S_1)$ with
$\mathcal{V}=v_c^4$ and $c\bar{c}_1({}^3S_1)+b\bar{b}_8({}^3S_1)$ with
$\mathcal{V}=v_b^4$, each of which has 6 Feynman diagrams. If $p_T$ is not
large enough, then these mixed contributions must dominate over 
$c\bar{c}_8({}^3S_1)+b\bar{b}_8({}^3S_1)$ by the enhancement factors
$1/v_b^4$ or $1/v_c^4$ while the double-fragmentation contribution
dominates at large $p_T$.

In addition, $c\bar{c}_8({}^3S_1)+b\bar{b}_8({}^{2s+1}L_J)$ and
$c\bar{c}_8({}^{2s+1}L_J)+b\bar{b}_8({}^3S_1)$, where ${}^{2s+1}L_J={}^1S_0$
or ${}^3P_J$, contribute to diagrams in 
Figs.~\ref{figpu}\,(b)--\ref{figpu}\,(d),   
with $\mathcal{V}= v_c^\alpha v_b^\beta$. Here, $\alpha,\,\beta=3$ for
${}^{2s+1}L_J={}^1S_0$ and 4 for ${}^3P_J$. The $\mathcal{V}$ is comparable
to that of the double-fragmentation, while they are suppressed at least by
either $v_c^3$ or $v_b^3$ in comparison with
$c\bar{c}_1({}^3S_1)+b\bar{b}_1({}^3S_1)$. Although the single-fragmentation
channel in Fig.~\ref{figpu}\,(c) may grow up at large $p_T$, that contribution
is dominated by the double-fragmentation [Fig.~\ref{figpu}\,(a)] by
a factor of $(m_c/p_T)^4$ or $(m_b/p_T)^4$. Hence, it is consistent
to ignore $c\bar{c}_8({}^3S_1)+b\bar{b}_8({}^{2s+1}L_J)$ and
$c\bar{c}_8({}^{2s+1}L_J)+b\bar{b}_8({}^3S_1)$  with ${}^{2s+1}L_J={}^1S_0$
or ${}^3P_J$ over the whole $p_T$ range.

As the last color-octet contribution, there are 
$c\bar{c}_8({}^{2s+1}L_J)+b\bar{b}_8({}^{2s'+1}L'_{J'})$, where 
${}^{2s+1}L_J,\,{}^{2s'+1}L'_{J'}={}^1S_0$ or ${}^3P_J$, whose typical
Feynman diagram is shown in Fig.~\ref{figpu}\,(d). In this case
$\mathcal{V}= v_c^\alpha v_b^\beta$ with $\alpha,\,\beta=3$  for
${}^{2s+1}L_J={}^1S_0$ and 4 for ${}^3P_J$. Thus the contributions are
suppressed compared to the color-singlet one. Because they do not have any
fragmentation contributions, they are dominated by the double-fragmentation
[Fig.~\ref{figpu}\,(a)] by $(m_c/p_T)^4(m_b/p_T)^4$ in the large-$p_T$ region.

So far, we have shown that, at LO in $\alpha_s$ and 
$v_{Q}$, it is sufficient to consider only
$c\bar{c}_8({}^3S_1)+b\bar{b}_8({}^3S_1)$,
$c\bar{c}_8({}^3S_1)+b\bar{b}_1({}^3S_1)$, and
$c\bar{c}_1({}^3S_1)+b\bar{b}_8({}^3S_1)$
to describe the $p_T$ spectrum of $p p \to J/\psi+\Upsilon + X$.
The parton-level differential cross sections for these processes are
available in Ref.~\cite{Ko:2010xy}.

Based on the formalism that we have described earlier, we are ready to carry
out the numerical calculation of the production rates for $J/\psi+\Upsilon+X$
at the LHC. The formula (\ref{xsec}) involves the factorization
of the PDF and the parton-level cross section
$d\hat{\sigma}$ with the factorization scale $\mu$. For the scale $\mu$, we
take the transverse mass $\mu=m_T=(4m_Q^2+p_T^2)^{1/2}$ with the heavy-quark
masses $m_c=1.5$ GeV and $m_b=4.7$ GeV. At LO in $\alpha_s$ on which
we are working, there are no additional hard jets and, therefore, $p_T$ and
$m_T$ are defined unambiguously. We employ the CTEQ6L parametrization
\cite{Pumplin:2002vw} for the PDF. We evaluate $\alpha_s$ by setting the
renormalization scale to be $m_T$ and use the next-to-leading-order formula
for the running coupling constant to be consistent with
Ref.~\cite{Pumplin:2002vw}.

In order to evaluate the production rate (\ref{xsec}), 
we have to know the values for the NRQCD
matrix elements $\langle O_n^{H}(^3S_1) \rangle$ for
$H=J/\psi$ and $\Upsilon$, where $n=1$ or 8.
The color-singlet matrix element $\langle O_1^H(^3S_1) \rangle$
is usually determined from the leptonic decay rate of $H$, which is
the most precisely measured value involving $H$.
We quote $\langle O_1^{J/\psi}(^3S_1) \rangle = 1.32\,{\rm GeV}^3$
\cite{psi-ME-1} and
$\langle O_1^{\Upsilon}(^3S_1)\rangle=9.21\,{\rm GeV}^3$ \cite{ups-ME-1}.
The color-octet matrix element $\langle O_8^{J/\psi} (^3S_1) \rangle$
has been fit to the $p_T$ spectrum of the inclusive prompt $J/\psi$
production rate at the Tevatron in the large-$p_T$ region.
The matrix element $\langle O_8^{\Upsilon} (^3S_1) \rangle$ has 
been fit \cite{Braaten:2000cm} to the Tevatron data and used for the
polarization analysis \cite{Braaten:2000gw}. Various determinations
of these matrix elements can be found, for example, in 
Ref.~\cite{Kramer:2001hh}. In this work, we use
$\langle O_8^{J/\psi} (^3S_1) \rangle = 3.9\times 10^{-3}$
\,${\rm GeV}^3$ \cite{Braaten:1999qk},
$\langle O_8^{\Upsilon} (^3S_1) \rangle = 1.5\times 10^{-1}$
\,${\rm GeV}^3$ \cite{Kramer:2001hh}.
\begin{figure}
\epsfig{file=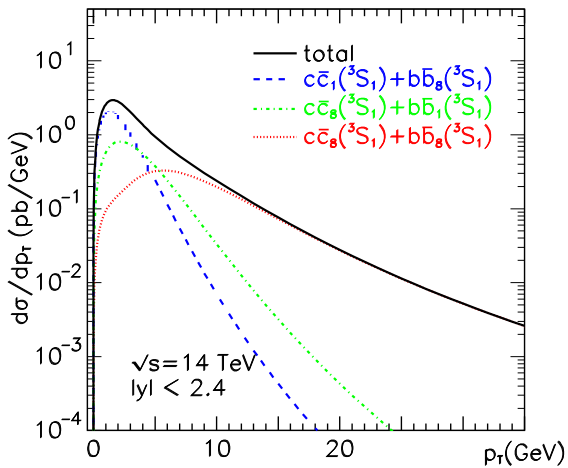,width=8cm}
\vspace*{-30pt}
\caption{\label{fig:psi-ups}%
The differential cross section $d\sigma/dp_T$ for
$pp\to J/\psi+\Upsilon+X$ at $\sqrt{s}=$14\,TeV 
in units of pb/GeV as a function of $p_T$ integrated over
the rapidity range $|y|<2.4$.
The solid, dashed, dashed-dotted, and dotted curves represent
the total, 
$c\bar{c}_1({}^3S_1)+b\bar{b}_8({}^3S_1)$, 
$c\bar{c}_8({}^3S_1)+b\bar{b}_1({}^3S_1)$, and
$c\bar{c}_8({}^3S_1)+b\bar{b}_8({}^3S_1)$
contributions, respectively.
}
\end{figure}

The prediction for $d\sigma/dp_T$ of $pp\to J/\psi+\Upsilon+X$
at $\sqrt{s}=$14\,TeV integrated over the rapidity range $|y|<2.4$ is shown
as the solid curve in Fig.~\ref{fig:psi-ups}. $d\sigma/dp_T$ vanishes
at $p_T=0$ and increases rapidly until it reaches the maximum value 
$d\sigma/dp_T|_{\rm max}=2.9$\,pb/GeV at $p_T=1.5$\,GeV.
Then it monotonically decays as $p_T$ increases.
Near $p_T=0$, $c\bar{c}_1({}^3S_1)+b\bar{b}_8({}^3S_1)$ dominates
and it is the largest for $p_T\lesssim 4$\,GeV. At large $p_T$,
the double-fragmentation contribution 
$c\bar{c}_8({}^3S_1)+b\bar{b}_8({}^3S_1)$ 
[Fig.~\ref{figpu}\,(a)] dominates and it is the largest
for $p_T \gtrsim$ 6\,GeV. In the remaining region,
$4\,\textrm{GeV}\lesssim p_T\lesssim$\, 6\,GeV, all of the three contributions
compete together.

As is mentioned earlier, we have ignored the color-singlet 
contribution in Fig.~\ref{figjusing} because it is suppressed compared 
to the color-octet one by a relative order of 
$\alpha_s^2$. In order to assure that this approximation is safe, 
we make a rough estimate of the color-singlet contribution to 
$pp\to J/\psi+\Upsilon+X$.
The corresponding parton processes consist of 3 types: Types A, B, and C are
shown in Fig.~\ref{figjusing}\,(a)--\ref{figjusing}\,(c), respectively.

We first classify the scaling behavior of the amplitude for the
double-fragmentation diagram shown in Fig.~\ref{figpu}\,(a).
The propagator for the exchanged gluon has the scaling $1/p_T^2$
and the gluon propagators attached to the $c\bar{c}$ and
$b\bar{b}$ pairs are of order $1/m_c^2$ and $1/m_b^2$, respectively.
The product of the two triple-gluon vertices must scale as
the typical momentum squared $p_T^2$. 
Therefore,
the resultant relative scaling of the double-fragmentation process is 
$v_c^4 v_b^4/(m_c^4m_b^4)$, where we have included the velocity-scaling
factor $\mathcal{V}=v_c^4v_b^4$.

In the case of type A,
there are four heavy-quark propagators and one gluon propagator
which have a large momentum transfer that accounts for the scaling
factor $1/p_T^6$. The phase space for the
$c\bar{c}_1+b\bar{b}_1+gg$ final state enhances the scaling
factor by an order of $p_T^4$ in comparison with
that for the two-body final state $c\bar{c}_{8}+b\bar{b}_{8}$
of the double-fragmentation process.
As a result, the type A process is scaled by $1/p_T^8$ 
compared to the double-fragmentation process. 
In a similar manner, the scaling factor for type B
is the same as that of type A.
In types A and B, there are extra hard jets in the final states.
Therefore, the color-octet contribution, which does not have hard
jets, is distinguished from these color-singlet contributions
that can simply be removed by imposing an appropriate veto.
Type C diagrams involve finite box diagrams with two virtual gluons
whose momentum must be of order the typical momentum transfer $p_T$.
A rough estimate of the scaling can be found by substituting a 
typical momentum transfer $p_T$ to the two gluon propagators,
four heavy-quark propagators, and the measure of the
loop momentum. The relative scaling of type C process is again $1/p_T^8$.

Therefore, the suppression factor for the color-singlet 
contribution relative to the color-octet double-fragmentation
process is $[\alpha_s/(4\pi)]^2[m_c/(v_{c}p_T)]^4\times$$[m_b/(v_{b}p_T)]^4$
where we have included the strong coupling suppression
factor $\alpha^2_s$ to the color-singlet contribution.
However, if $p_T$ is small, then the factors $p_T$ in this
scaling should be of order $m_Q$ and, therefore, the suppression
factor becomes roughly $\alpha_s^2/[(4\pi)^2 v_{b}^4]$ in comparison with
$c\bar{c}_1({}^3S_1)+b\bar{b}_8({}^3S_1)$, which dominates over the
double-fragmentation contribution as $p_T\to 0$. This factor is of order
$10^{-2}$ if we assume $v_b^2\sim 0.1$ and we use the renormalization
scale of order $m_b$. According to this rough estimate, 
we expect that our prediction may not be contaminated by
an order-$\alpha_s^6$ color-singlet contribution even at small $p_T$. 
If we introduce a lower $p_T$ cut $p_T \gtrsim 5$\, GeV, where the
double-fragmentation rises up, then the color-octet
contribution becomes more significant.
In addition, by imposing a veto 
that the final state must not include hard jets, one can remove
the color-singlet contribution of types A and B,
which are potential sources of a large background.

The kinematic enhancement of the double gluon fragmentation can
also be applied to the color-singlet processes of order $\alpha_s^8$
through the subprocesses $gg,\, q\bar{q} \to gg$ followed by the
fragmentations of the two final-state gluons into 
$c\bar{c}_1({}^3S_1)+ gg$ and $b\bar{b}_1({}^3S_1)+gg$, respectively.
This process is suppressed compared to the color-octet double
gluon fragmentation $g\to c\bar{c}_8({}^3S_1)$ and
$g\to b\bar{b}_8({}^3S_1)$ that we have calculated in this paper
by a factor of order 
$\alpha_s^4 / [(4\pi)^4 v_c^4 v_b^4] \sim 7 \times 10^{-5}$
with an assumption $\alpha_s=0.2$ at threshold with the scale
$\mu=2m_Q$.\footnote{
Our order-of-magnitude estimate relies somewhat on the extra factors
of $(4\pi)$, which are from the phase space factor or loop integration.
It is well known that the inclusion of the factor $(4\pi)$ improves the
naive estimates, for example, in the muon decay rate using the Fermi's
four-fermion interaction Lagrangian for charged weak interaction.
Our naive estimates may break down if there are large enhancement 
factors in the color-singlet channels. An explicit calculation of
the complete color-singlet contribution may reveal if our rough power
counting is indeed reasonable.}
In addition, one can consider 
the mixed double gluon fragmentation of order $\alpha_s^6$
through the subprocesses $gg,\, q\bar{q} \to gg$ followed by the
fragmentations of the two final-state gluons into 
$c\bar{c}_8(b\bar{b}_8)({}^3S_1)$ and
$b\bar{b}_1(c\bar{c}_1)({}^3S_1)+ gg$, respectively.
These mixed double-fragmentation contributions are also 
suppressed relative to the color-octet counterpart by factors of  
$\alpha_s^2 / [(4\pi)^2 v_b^4] \sim 3 \times 10^{-2}$ and
$\alpha_s^2 / [(4\pi)^2 v_c^4] \sim 3 \times 10^{-3}$
for the $c\bar{c}_8+b\bar{b}_1$ and 
$b\bar{b}_8+c\bar{c}_1$ fragmentations, respectively.\footnote{
If we make use of the fragmentation probability at threshold
given in Ref.~\cite{Braaten:1995cj}, we can make rough estimates on
the ratios of the double-fragmentation processes
$c\bar{c}_1+b\bar{b}_1$,
$c\bar{c}_8+b\bar{b}_1$, and
$b\bar{b}_8+c\bar{c}_1$ compared
to the $c\bar{c}_8+b\bar{b}_8$ fragmentation as
$\sim 2\times 10^{-4}$, $\sim 5\times 10^{-3}$, 
and $\sim 4\times 10^{-2}$, respectively.
These values are consistent with the scaling factors
listed above.
} In summary, the mixed double-fragmentation contributions may occupy
only a few percents of the color-octet double fragmentation and the
pure color-singlet double-fragmentation contribution seems to be well
suppressed compared to the color-octet double-fragmentation
contribution.

Based on the NRQCD factorization formalism, we have computed the $p_T$
spectrum $d\sigma/dp_T$ and the total cross section 
$\sigma_{\textrm{tot}}$ for $pp\to J/\psi+\Upsilon+X$.
The short-distance coefficients were computed at
LO in $\alpha_s$ and $v_{Q}$. $\sigma_{\textrm{tot}}$
integrated over the rapidity and the transverse-momentum range
$|y|<2.4$ and $p_T<30$\,GeV is predicted to be
$\sigma_{\textrm{tot}}=$ 13\,pb at $\sqrt{s}=14$\,TeV. 
Because the production rate is subject to
the numerical values for $\langle O_8^{J/\psi}({}^3S_1)\rangle$ 
and $\langle O_8^{\Upsilon}({}^3S_1)\rangle$, 
the forthcoming empirical rate may provide information whether 
the values were reasonable or not.

The feature of this LO process that the color-singlet
channel is suppressed provides us with a clean probe to
the color-octet mechanism.
In comparison with the inclusive single-quarkonium production
at the hadron colliders, the suppression of the color-singlet channel
is more significant by a relative factor of 
$[\alpha_s/(4\pi)]^2 [m_Q/(v_Q p_T)]^4$ at large $p_T$.
The $p_T$ spectrum is dominated by $c\bar{c}_1({}^3S_1)+b\bar{b}_8({}^3S_1)$
and $c\bar{c}_8({}^3S_1)+b\bar{b}_8({}^3S_1)$ channels at the 
low- and large-$p_T$
region, respectively, and the $c\bar{c}_8({}^3S_1)+b\bar{b}_1({}^3S_1)$
contribution is comparable to those two listed above only around the region 
$4\,\textrm{GeV}\lesssim p_T\lesssim 6\,\textrm{GeV}$. 
Assuming the integrated luminosity 
$\sim 100  \,\textrm{fb}^{-1}$ at $\sqrt{s}=$14\,TeV and
considering the branching fractions $B[J/\psi\to \mu^+\mu^-]=5.93\,$\% 
and $B[\Upsilon \to \mu^+\mu^-]=2.48\,$\%~\cite{Amsler:2008zzb}, 
we expect that approximately 1900, 520, and 160 events can be observed 
by tagging muon pairs under the cuts $p_T\ge$ 0, 5, and 10\,GeV, 
respectively, at the LHC. If one can improve the
acceptances for $J/\psi$ and $\Upsilon$ by extensive Monte Carlo
studies of final-muon pairs, then the observation of the events 
can be quite promising in the near future.
Inclusion of $e^+e^-$ decay modes of $J/\psi$ and $\Upsilon$
may increase the number of events by a factor of 4.
One can also improve the prediction by including 
the subprocess $q\bar{q}\to J/\psi + \Upsilon$ via two-gluon
exchange that is neglected in this work.
If the forthcoming measured rate is significantly less than our
prediction, then it may be an indication that the current values for
the octet matrix elements are overestimated.

We consider possible sources that may contaminate our predictions.
In this work, we have computed the rate for the direct $J/\psi$'s rather
than prompt $J/\psi$'s that include feeddowns from higher resonances
and that exclude those from the $B$-meson decay. While
nonprompt signals from $B$ decays can be eliminated with the aid of 
silicon vertex detectors, the separation of the feeddowns 
from higher resonances is not easy.
Therefore, the simultaneous determination of
the color-octet NRQCD matrix elements
$\langle O_8^{J/\psi}({}^3S_1)\rangle$ and
$\langle O_8^{\Upsilon}({}^3S_1)\rangle$ using this process may be nontrivial.
However, the dominance of the $c\bar{c}_8(^3S_1)+b\bar{b}_8(^3S_1)$
contribution may still be true, even after including the feeddowns
from higher resonances, especially at large $p_T$ due to the
double-fragmentation process.
Once the $p_T$ distribution for the $J/\psi+\Upsilon$ production
is measured, it may provide us with significant constraints 
on the product of the two matrix elements.

We note that the total rate and the $p_T$ spectrum for the inclusive
single-quarkonium production at hadron colliders suffer from large
next-to-leading-order corrections in $\alpha_s$. In particular,
the next-to-leading-order corrections to the color-singlet contribution 
to the inclusive $J/\psi$ production enhance the rate 
by an order of magnitude~\cite{Campbell:2007ws,Gong:2008sn,Artoisenet:2008fc}.
Therefore, it is natural to worry that the inclusion of order-$\alpha_s^6$
contributions to this process may modify the $p_T$ spectrum significantly.
However, in the case of $J/\psi+\Upsilon$
production, the suppression factor of the color-singlet channel
to the color-octet one is $\alpha_s^2/p_T^8$ which is significantly
smaller than the corresponding factor $1/p_T^4$ for the single-quarkonium
production, in which the color-singlet and color-octet
contributions are of the same order ($\alpha_s^3$) at LO.
We anticipate that our arguments can be tested quantitatively by measurements
and by explicit calculations of complete order-$\alpha_s^6$ contributions
to the color-singlet channel in near future.
Once the suppression of the color-singlet channel is confirmed at that order,
then the measurement of the $p_T$ spectrum of $J/\psi+\Upsilon$ at the LHC
may play an important role to determine the color-octet NRQCD matrix elements
$\langle O_8^{J/\psi}({}^3S_1)\rangle$ and
$\langle O_8^{\Upsilon}({}^3S_1)\rangle$. Even in the worst case, that 
one could not observe the $p_T$ spectrum at the desired level, 
at least one can obtain strong upper bounds to both NRQCD matrix elements,
that were not easy to extract empirically in other experiments.
Therefore, the forthcoming measurement may provide useful information
that can be used to understand the mechanism of the decade-old puzzle 
for the polarization of prompt $J/\psi$ at the Tevatron.

In summary, we have computed the production rate and $p_T$ distribution
for $pp\to J/\psi + \Upsilon + X$ at the LHC in the framework of the NRQCD
factorization formalism at leading order in $\alpha_s$.
Unlike most inclusive single-quarkonium-production processes,
in which at least three independent color-octet NRQCD matrix elements
$\langle O_8^{H}({}^3S_1)\rangle$,
$\langle O_8^{H}({}^1S_0)\rangle$, and
$\langle O_8^{H}({}^3P_J)\rangle$ involve,
this process depends dominantly on two matrix elements
$\langle O_8^{J/\psi}({}^3S_1)\rangle$ and 
$\langle O_8^{\Upsilon}({}^3S_1)\rangle$ 
at LO in $\alpha_s$, especially at large $p_T$.
If the corresponding measured rate is significantly less than the
prediction presented in this work,
it may imply that current values for the octet matrix elements 
are overestimated. The forthcoming measurement of $p_T$ distribution
may probe to the color-octet mechanism in quarkonium production,
by imposing strong constraints to these NRQCD matrix elements.

\vspace{0.5cm}

\begin{acknowledgments}
The authors would like thank Kuang-Ta Chao and Rong Li
for their useful comments. They also thank Suyong Choi for providing
valuable information regarding four-muon analysis at the CMS experiment.
A part of the work (P.K.) was done at Aspen Center for Physics during the
summer workshop in 2010.
The work of P.K. was supported in part by the National Research 
Foundation (NRF) through the Korea Neutrino Research Center (KNRC) 
at Seoul National University. 
This work was supported
by the Basic Science Research Program through the NRF of Korea funded by the
MEST under Contracts No. 2009-0072689 (C.Y.) and No. 2010-0000144 (J.L.).
\end{acknowledgments}




\end{document}